# Predicting SLA Violations in Real Time using Online Machine Learning

Jawwad Ahmed[1], Andreas Johnsson[1], Rerngvit Yanggratoke[2], John Ardelius[3], Christofer Flinta[1], Rolf Stadler[2]

[1] Ericsson Research, Sweden, Email:{jawwad.ahmed, andreas.a.johnsson, christofer.flinta}@ericsson.com
[2] ACCESS Linnaeus Center, KTH Royal Institute of Technology, Sweden, Email: {rerngvit,stadler}@kth.se
[3] Swedish Institute of Computer Science (SICS), Sweden, Email:john@sics.se

*Abstract*— Detecting faults and SLA violations in a timely manner is critical for telecom providers, in order to avoid loss in business, revenue and reputation. At the same time predicting SLA violations for user services in telecom environments is difficult, due to time-varying user demands and infrastructure load conditions.

In this paper, we propose a service-agnostic online learning approach, whereby the behavior of the system is learned on the fly, in order to predict client-side SLA violations. The approach uses device-level metrics, which are collected in a streaming fashion on the server side.

Our results show that the approach can produce highly accurate predictions (>90% classification accuracy and < 10% false alarm rate) in scenarios where SLA violations are predicted for a video-on-demand service under changing load patterns. The paper also highlight the limitations of traditional offline learning methods, which perform significantly worse in many of the considered scenarios.

## I. Introduction

Next generation telecom services will execute on the telecom cloud, which combine the flexibility of today's computing clouds with the service quality of telecom systems. Real-time service assurance will become an integral part in transforming the general and flexible cloud into a robust and highly available cloud that can ensure low latency and agreed service quality to its customers.

A service assurance system for telecom services must be able to detect and preferably also predict problems that may violate the agreed service quality. This is a complex task already in legacy systems and will become even more challenging when executing the services in the cloud. Further, the service assurance system must be able to remedy, in real time, these problems once detected.

One promising approach to service assurance is based on machine learning, where the service quality and behavior is learned from observations of the system. The ambition is to do automated real-time predictions of the service quality in order to execute mitigation actions in a proactive manner.

Machine learning has been used in the past to build prediction models for service quality assurance. For example, predicting user application quality-of-service (QoS) parameters [1][2], quality-of-experience (QoE) [3], and anomalies for complex telecom environments [4]. The key idea is to use sample data for training a statistical model, which is then used for predictions of unseen data. This work contrasts from previous work in that it makes service quality predictions based on low-level device statistics, and that the model is allowed to learn new concepts from a stream of examples in real time.

This paper presents results from experiments using online machine learning where the aim is to predict service quality in terms of SLA fulfillment from system-level device statistics, see illustration in Figure 1. Specifically, we experiment with a video-on-demand service (VoD) and try to predict, on the server side, whether the agreed service quality on the client side is violated or not (i.e., a binary classification problem). The service quality is dictated by a service level agreement, SLA, signed by the provider and the user. The online approach allows the prediction model to learn the behavior of the system as time progresses. In other words, the understanding of the system behavior is gradually improved as new data becomes available. This is referred to as incremental learning and it eliminates the need for an explicit offline phase to fully train the model before it can be used.

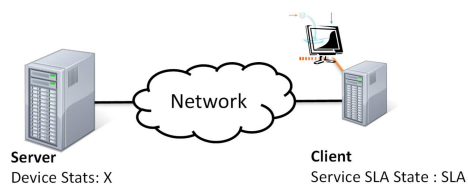

Fig. 1. Sceneraio setup.

Predicting the SLA fulfillment is one important tool for delivering high-quality services. The provider can take timely and faster actions based on predictive analysis compared to traditional slower and error prone customer support services.

In our previous work [1] we show that it is feasible to predict VoD service quality from system-level statistics. This paper extends our previous work by identifying limitations with the offline methodology and instead proposes and evaluates an online approach. Further, our previous work focuses on the regression problem while this paper specifically studies the SLA fulfillment classification problem.

In summary, this paper has the following main contributions:

- introduces an application agnostic approach for real-time prediction of SLA violations based on online machine learning;
- shows that accurate prediction of SLA violation can be achieved using online learning methods for a VoD service scenario;
- demonstrates that online learning methods significantly outperform offline learning methods in scenarios with concept drift.

The rest of this paper is organized as follows: Section II defines the problem. Section III discusses machine learning from streaming data. Section IV describes feature space and service quality metrics. Section V details the testbed setup. An extensive performance evaluation is conducted and discussed in Section VI. Conclusions are drawn in Section VII.

## II. PROBLEM SETTING

The envisioned system consists of a set of servers connected to another set of clients over a network. Each client can access a VoD service running on the server side. Note that a similar setup was investigated in our previous work [1][5]. However, compared to our previous work this setup assumes a stream of learning examples from at least one client in order to build an online service quality prediction model for the clients in real time. The experimental setup is visualized in Figure 2.

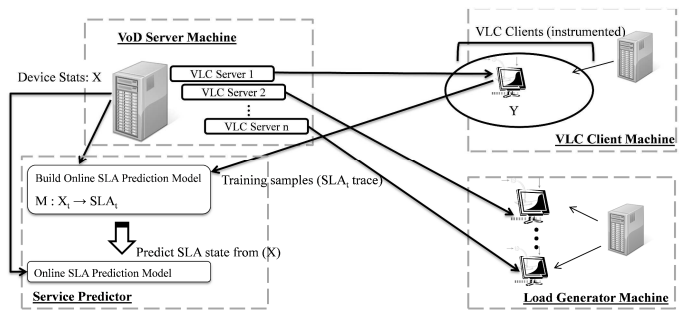

Fig. 2. Test-bed setup.

Device statistics $X$ are collected at the server while the service is operational. In this setting, device statistics refer to system metrics on the operating-system level such as CPU utilization, the number of running processes, the rate of context switches and free memory. In contrast, service-level metrics $Y$ on the client side refer to statistics on the application level such as the video frame rate. The metrics $X$ and $Y$ are fed in real time to the *Service Predictor* module.

The metrics $X$ and $Y$ evolve over time, influenced by the load on the server, operating system, and application load dynamics. Assuming a global clock, that can be read on both the client and the server, we can model the evolution of the metrics $X$ and $Y$ as time series $\{X_t\}_t$, $\{Y_t\}_t$ and $\{(X_t, Y_t)\}_t$. The assumption in [1] was that the metrics are drawn from a stationary and unknown distribution $D$. Under this assumption it was possible to create a prediction model $M : X \rightarrow Y$ using offline machine learning techniques.

However, the assumption of a stationary distribution is limiting and does not always hold in real-world scenarios. The underlying problem is often referred to as a concept drift. That is, instead of being stationary the distribution $D$ evolves over time: $\{D_1, D_2, ..., D_t\}$. It implies the prediction system cannot rely on an initial model M throughout the service lifecycle without any changes. Rather, the model M has to be updated as new samples arrive. The prediction system will end up with an evolution of models $\{M_t, t=1..T\}$. This puts new demands on the actual learning but also on how to build the actual system.

Note that we do not discuss the true origin or cause of the drifting distribution $D$; rather we observe problems when learning from one trace and predicting on a second trace and try to overcome that with online learning.

Assume that a service level agreement (SLA) has been defined and computed for the client side service metrics. The SLA at a given time can either be in a 'violated' or 'not violated' state based on the service metric values. That is, for each service metric $Y_t$ at time t we can compute an $SLA_t$ value.

The objective in this paper is to predict application level SLA conformance for clients based on $X$, under the assumption that the underlying distribution $D$ may change over time.

Further note that we assume that the distribution D does not depend on the network or the client. In practical terms, this means that the network is lightly loaded. We are aware that such assumptions may not hold for real systems, and we plan on relaxing them in the future. Previous research has also shown how end-to-end network measurements can be used to predict client-side metrics [2]; therefore, including this aspect would not necessarily add to the contributions of this work. Rather, we believe that this work complement already existing work on performance prediction in the management area.

## III. LEARNING FROM HIGH SPEED DATA STREAMS

Data stream processing essentially differs from traditional data processing in that the statistical calculations need to be made using an infinite number of continuously arriving samples [6]. This restricts the number of available models to those that can be efficiently trained using a single pass over the data. The practical rational behind this is that data is generated with high velocity and in large volumes, limiting the storage and multi-pass processing.

Machine learning refers to the task of predicting unseen samples based on the currently available examples using statistical learning techniques. Predictions can broadly be categorized into a classification problem or regression problem [7]. In this paper we limit our study to classification problem, which contrasts to our previous work in [1] which studied a similar setup but for regression. Suitable online models for predictions should exhibit low prediction and learning latency, low overhead, and preferably also be interpretable (often refered to as a white box solution).

An inherent problem of real world data not always considered in theory is that of concept drift where the learned relation between independent and dependent variables changes over time. In terms of notations in this paper it simply means that the underlying and unknown distribution D, defined in the previous section, changes over time. To cope with such situations models need to be adaptive. This further motivates our preference for online models since models learned offline can quickly become outdated with degraded accuracy when handling evolving data streams [8].

Online learning example.

Below we briefly describe the online machine learning methods used in this work.

Logistic regression [7] is a common method for solving classification problems. It specifies the log-odds posterior probability for the classes and takes positive or negative real numbers as input, while output values are either zero or one. It can hence be interpreted as the posterior probability of drawing a sample from the respective class. The coefficients of the model can efficiently be trained using stochastic gradient decent (SGD) which makes it suitable for online applications. In this paper we use Gradient Ascent Logistic Regression as an online methodology.

Another family of classifiers is decision trees, where the feature space is split into a set of regions based on one or more homogeneity criteria recursively until there is no more significant gain in terms of homogeneity of the members of each class. Some commonly used implementations of the decision trees are C4.5 [9] and CART [10]. The benefit of decision trees over linear methods is that they perform well even when the inherent relationship between the predictors and the response variable is non-linear. A non-parametric property means that no assumptions are made on the form or shape of the true underlying function. The cost of making a decision is dependent on the height of the tree which is logarithmic in the size of tree.

In this paper we specifically apply Very Fast Decision Tree (VFDT or Hoeffding Tree) [11], an incremental learning algorithm for creating a decision tree in constant time per training sample (but proportional to the number of features). The algorithm applies the Hoeffding bound, which ensures that the constructed tree asymptotically does not differ from a tree constructed using conventional batch algorithm. The algorithm has a low memory and computational complexity requirements.

In general, for both offline and online versions, decision trees are flexible and have low bias but may lead to high variance in predictions in many cases. However, studies have shown that their effectiveness can be significantly increased when used as a base component for the ensembles such as bagging or random forest [7]. Our previous study [1] confirms these findings (for regression) where we found random forest to be the most consistent performer in different offline learning scenarios.

Due to this fact we also use a third online algorithm, the Online Accuracy Update Ensemble (OAUE) which is an ensemble method that combines batch learning with incremental classifiers to learn from data streams under concept-drift scenarios [12]. The algorithm relies on setting classifier weights according to some quality measure for predicting on a sliding window of training samples.

Most online methods handle concept drift by slowly learning the new relation with arrival of new samples. The new samples adapt the model and hence the model accuracy will increase over time. Concept drifts can also be handled explicitly by coupling any of the three learning algorithms mentioned above with change detecting algorithms such as Adwin [13] or Probabilistic Adaptive Window [14]. Here, we limit the study to the implicit concept drift handling, and rely on the online method to learn a new concept over time.

Actual selection of online machine learning algorithm and parameter tuning requires careful performance analysis of the algorithm characteristics with extensive experimentation in the relevant domain area. The aim of this paper is to show that online methods can tackle the problem of concept drifts, using the above algorithms as examples, rather than providing an exhaustive investigation to find the best algorithm.

IV. DEVICE STATISTICS AND SLA PREDICTION

A. Device Statistics

Device statistics, denoted as a feature matrix $X$, are defined as metrics obtained from the Linux kernel at the server side. Specifically, we utilize the System Activity Report (SAR), an open source Linux tool [15] reading data through the Linux process file system, to construct $X_{SAR}$. Metrics are computed over a configurable interval which is set to 1 second. Based on $X_{SAR}$ we construct $X$ by excluding non-numerical features and performing manual feature selection and aggregation operations. The resulting $X$ contains 21 features (see Table I).

This feature set is similar to the one that is gathered by various other monitoring tools such as Ceilometer for the compute nodes in OpenStack [16]. A small aggregated feature space containing only the key system metrics is critical for low-latency driven online learning on high speed data streams.

Note also that this reduced feature set approach is different from our previous work [1] where the feature space was comprised of raw kernel level and SAR metrics with limited preprocessing. There the size of the feature set was not an issue due to the offline methodology.

B. Service Quality Metrics and SLA Prediction Approach

A service level agreement (SLA) is the contract between a provider and its customers. It describes the promised quality-of-service, responsibilities and allowed divergences in terms of various metrics. An SLA may comprise of a set of service level objectives (SLO) which provide a way of measuring the service quality and associated thresholds.

In this paper, we specifically study a VoD service, realized by instrumented VLC [17] media player software, where the video is streamed from the server to the clients (see Figure 2).

The service-quality metrics considered are measured on a client. During an experiment we capture the following two metrics which are used for the SLA computation.

1. *Video frame rate (frames/sec): the number of displayed video frames per second (FPS);*

2. *Audio buffer rate (buffers/sec): the number of played audio buffers per second (ABS);*

For the VoD service we construct an SLA based on two SLOs; namely the video frame rate denoted $SLO^{FPS}$ and the audio buffer rate denoted $SLO^{ABS}$. The two SLOs are either in a violated state or in a non-violated state. The SLA is in violated state if any of these SLOs is violated:

$$SLA\ violated = (SLO^{FPS} = violated) \vee (SLO^{ABS} = violated) \quad (1)$$

where the SLO is violated if the corresponding metric value drops below a certain threshold.

Note that while in the evaluation of this paper we only focus on $SLO^{FPS}$ since it was shown to be more difficult to predict audio buffer rate using regression models in general, see previous results in [1].

TABLE I.    SYSTEM METRICS FOR CONSTRUCTING X FEATURE SPACE

| CPU [%] | Memory/Swap [Kb] | I/O Trans. per sec | Block I/O per sec | Proc Stats per sec | Network Stats per sec |
|---|---|---|---|---|---|
| CPU Idle | Mem Used | Read Trans. | Bock Reads | New Processes | Received Packets |
| CPU User | Committed Mem | Write Trans. | Block Writes | Context Switches | Transmitted Packets |
| CPU System | Swap Used | Bytes Read | | | Received Data (KB) |
| CPU_IO Wait | Cached Swap | Bytes Written | | | Transmitted Data (KB) |
| | | | | | Interface Unitization (%) |

## V. TESTBED AND EXPERIMENTATION

This section describes the hardware and software setup, methodology to perform experiments, generate load and how to collect sample traces for model training.

### A. The testbed

The testbed used for experimentation includes 60 rack based servers that are interconnected by Ethernet switches. The machine configuration is Dell PowerEdge R715 2U rack servers, 64 GB RAM, two 12-core AMD Opteron processors, 500 GB hard disk, and a 1Gb Ethernet network controller.

The basic setup for experimentation is shown in Figure 2 and includes three physical machines, namely, a server that provides the VoD service, a client that runs a VoD session, and a load generator that creates the aggregate demand of a set of VoD clients. All machines run Ubuntu 12.04 LTS, and their clocks are synchronized through NTP [18].

The server machine runs one or more VLC servers (version 2.1.3). Each VLC server is configured for VoD service. It transcodes the video and audio data, and streams the results over the network to the VoD clients. The sensor on the server machine periodically reads out and stores the vector $X_{SAR}$ (SAR version 10.0.3) and a timestamp. Then $X_{SAR}$ is filtered and aggregated into an X trace file as described above.

The server machine is populated with the ten most popular YouTube videos from 2013. The client machine runs an instrumented VLC client, and a sensor extracts service-level events from it. At the start of every second, the sensor collects the events from the last second, computes the Y metrics, and writes them to the local Y trace file, together with a timestamp. The load generator machine dynamically spawns and terminates VLC clients, depending on the specific load pattern that is being executed during an experiment.

Figure 2 also depicts how the samples can be distributed over the network for online learning. Training samples must be transferred from at least one client and also from the VoD Server Machine to the *Service Predictor*. Online predictions of the SLA are then performed.

At the beginning of a run, the VLC client sends a request for playing a specific video to a VLC server, which is selected uniformly at random among a pool of VLC servers. Once the video has played out, the VLC client sends a new request for the same video to a random VLC server. Also, at the beginning of the run, the load generator starts sending requests according to the selected load pattern, and the sensors on the server and client machines are initialized. Depending on the selected load pattern the utilization of the server machine can vary significantly.

### B. The data set

As discussed in Section II, the main focus of this paper is to explore scenarios with a real or apparent drifting distribution $D$. The following paragraphs describe how a set of load traces are created.

Load traces are created via a load pattern generator that dynamically controls the number of currently active VoD sessions by spawning and terminating VLC clients (see Figure 2). The following load traces [19] were generated:

*Periodic T1 load trace:* the load generator starts clients following a Poisson process with an arrival rate that starts at 30 clients/minute and changes using a sinusoid function with a period of 60 minutes and amplitude of 20 clients/minute. The load generator terminates a client after an exponentially distributed holding time with average of 1 minute.

*Periodic T2 load trace:* the same configuration parameters are used as for the Periodic T1, but this trace was executed at different point in time and with a slightly smaller duration of the total experiment run time.

*Flash Crowd load trace:* the load generator starts and terminates clients according to a flash-crowd model described in [20]. The creation of clients follows a Poisson process with an arrival rate that starts at 5 clients/minute and peaks at flash events, which are randomly created at a rate of 10 events/hour. At each flash event, the arrival rate linearly increases 10-fold to 50 clients/minute within one minute; it sustains this level for one minute, and then linearly decreases to 5 clients/minute within 4 minutes. The load generator terminates a client after an exponentially distributed holding time, 1 minute average.

The two periodic load traces are used to investigate the impact on model accuracy due to time differences between traces while the flash crowd trace is used to study accuracy originating from changing load scenario.

## VI. EVALUATION

In this section, we describe the evaluation methodology, metrics for method comparison as well as a detailed set of results obtained both using offline and online methods.

In this evaluation, we address the following research questions:

1. What is the classification accuracy of an offline learning method applied to a specific load trace? How accurate is a model trained on one trace and evaluated on a different trace?

2. How does the classification accuracy of an offline method differ from that of an online method for traces with and without concept drift?

### A. Evaluation Methodology

The main parts of the performance evaluation framework have been implemented in R. The RMOA library [21] has been used by interfacing R with the MOA stream mining system [22] to execute online learning. One baseline online method, gradient ascent logistic regression was evaluated in MATLAB.

The validation set approach [7] is used for evaluation of offline methods; 70% of samples in training set, 30% testing set. Note that offline models do not evolve with time. Hence, for evaluating concept drift scenarios, the offline model is first trained with data from one load trace and then tested on another load trace. This way it is possible to assess the impact of the concept drift on the model prediction accuracy.

The *test-then-train* approach with chunks is used for evaluating online learning methods [13]. The basic operation of this approach is to test and then train the model based on a stream of samples in a chunk. In this paper the chunk size is set to 10 samples. Further a bootstrap chunk of 500 samples is used to do the initial training. Other chunk and bootstrap sizes were also evaluated, but no significant difference in prediction accuracy was observed. However, note that a lower number than 500 bootstrap samples caused stability problems for ensemble learning methods in MOA.

To compare models we use the following traditional performance metrics: classification accuracy (CA), balanced accuracy (BA), true positive rate (TPR), true negative rate (TNR) and false alarm rate (FAR), calculated as follows:

$$CA := \frac{True\ Positives + True\ Negatives}{Total\ Test\ Samples} \quad (2)$$

$$BA := \frac{TPR + TNR}{2} \quad (3)$$

$$TPR := \frac{True\ Positives}{True\ Positives + False\ Negatives} \quad (4)$$

$$TNR := \frac{True\ Negatives}{True\ Negatives + False\ Positives} \quad (5)$$

$$FAR := \frac{False\ Negtives}{False\ Negatives + True\ Positives} \quad (6)$$

The classification accuracy metrics are calculated over either (1) an interval increasing proportional to the number of samples and (2) a sliding window with a fixed size. The interval accuracy shows the asymptotic behavior while the sliding window provides insights on the temporal aspects.

Three different representative offline methods have been used to evaluate offline prediction accuracy: batch logistic regression (standard R), decision trees ('Tree' package of R) and random forest ('RandomForest' package of R).

For the online counterparts of the offline methods we selected gradient ascent logistic regression as a substitute for batch logistic regression with a learning rate of 0.01. The number of gradient ascent iteration per chunk was set to 100.

The online version of decision trees used is the popular Hoeffding Tree implementation VFDT [11]. For VFDT we used the following parameter settings: adaptive Naive Bayes leaf predictions with a grace period of $n_{min}= 100$, split confidence $\delta = 0.01$ and tie-threshold $\tau_i = 0.05$.

For online ensemble we use the Online Accuracy Updated Ensemble (OAUE) algorithm [12] and Hoeffding Tree is used as the base learner, with a maximum of 100 trees.

### B. SLO thresholds

To determine SLO thresholds, as discussed above, certain knowledge about the service metrics is required. Figure 3 shows a histogram of the video frame rate and audio buffer rate. Note that both audio and video are bimodal. Based on this we set the threshold $SLO^{FPS} = 20$ and $SLO^{ABS} = 20$. Both these values divide the two distributions into two regions with one mode in each region (violated and non-violated).

To validate the two regions we manually observed the video quality for frame rates above and below 20 FPS. A frame rate below 20 delivers a choppy video experience, and thus the service quality was considered as violated in this case. Note that the impact of $SLO^{ABS}$ threshold is currently being investigated.

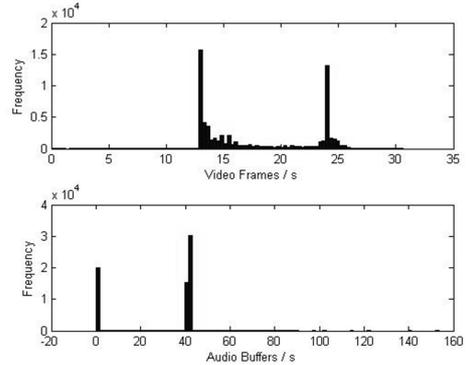

Fig. 3. Video frame rate and audio buffer rate under variable load scenarios.

### C. Evaluation Results

To answer the first research question we present experimental results from conducting offline learning for different load traces. The results are shown in Table II. The best results are highlighted for each load trace.

TABLE II. PERFORMANCE FOR OFFLINE METHODS

| Classification Method | Load Trace | CA | BA | TPR | FAR |
|---|---|---|---|---|---|
| Logistic Regression | Periodic T1 | 0.858 | 0.836 | 0.897 | 0.106 |
| | Periodic T2 | 0.873 | 0.857 | 0.901 | 0.087 |
| | Flashcrowd | 0.906 | 0.905 | 0.894 | 0.101 |
| Decision Tree | Periodic T1 | 0.890 | 0.860 | 0.944 | 0.100 |
| | Periodic T2 | 0.860 | 0.831 | 0.909 | 0.111 |
| | Flashcrowd | 0.899 | 0.900 | 0.902 | 0.121 |
| Random Forest | Periodic T1 | **0.913** | **0.886** | **0.961** | **0.084** |
| | Periodic T2 | **0.937** | **0.921** | **0.965** | **0.055** |
| | Flashcrowd | **0.910** | **0.910** | **0.903** | **0.100** |

TABLE III. PERFORMANCE FOR ONLINE METHODS

| Classification Method | Load Trace | CA | BA | TPR | FAR |
|---|---|---|---|---|---|
| Gradient Ascent Logistic Regression | Periodic T1 | 0.901 | 0.879 | 0.817 | 0.130 |
| | Periodic T2 | 0.928 | 0.915 | 0.877 | 0.103 |
| | Flashcrowd | 0.896 | 0.896 | 0.900 | 0.101 |
| Hoeffding Tree | Periodic T1 | 0.881 | 0.850 | 0.938 | 0.107 |
| | Periodic T2 | 0.873 | 0.834 | 0.940 | 0.118 |
| | Flashcrowd | 0.885 | 0.886 | 0.894 | 0.134 |
| Online Accuracy Updated Ensemble | Periodic T1 | **0.913** | **0.890** | **0.954** | **0.080** |
| | Periodic T2 | **0.933** | **0.916** | **0.962** | **0.059** |
| | Flashcrowd | **0.910** | **0.901** | **0.901** | **0.100** |

Observe the high classification accuracy for all traces, and also note that the balanced accuracy numbers are close to the classification accuracy. This indicates that the classifiers perform well when predicting both negative and positive responses (i.e. violated and non-violated SLA).

Random forest has the highest performance (CA above 0.9) on all three load traces. However, using simple logistic regression and decision tree methods also provided decent performance (CA above 0.85). The results in Table II can be considered a baseline for a stationary scenario that the online learning can be compared to.

TABLE IV. RANDOM FOREST PERFORMANCE IN CONCEPT DRIFT SCENARIOS

| Load Trace (Training) | Load Trace (Testing) | CA | BA | TPR | FAR |
|---|---|---|---|---|---|
| Periodic T1 | Periodic T2 | 0.464 | 0.599 | 0.232 | 0.062 |
| | Flashcrowd | 0.904 | 0.904 | 0.893 | 0.103 |
| Periodic T2 | Periodic T1 | 0.772 | 0.671 | 0.952 | 0.232 |
| | Flashcrowd | 0.812 | 0.821 | 0.933 | 0.270 |
| Flashcrowd | Periodic T1 | 0.855 | 0.842 | 0.878 | 0.095 |
| | Periodic T2 | 0.677 | 0.742 | 0.564 | 0.061 |

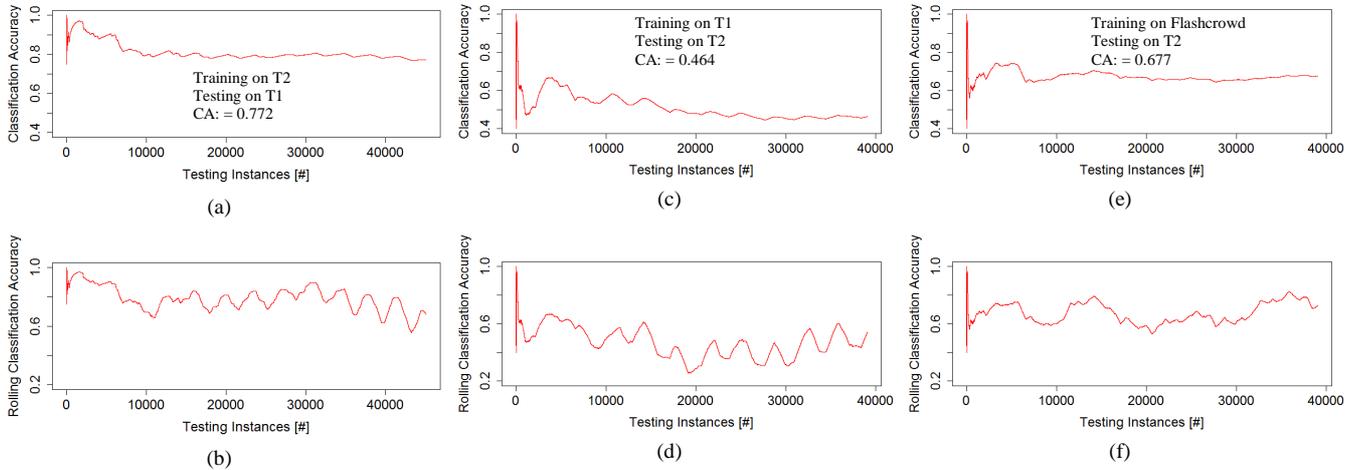

Fig. 4. Asymptotic classification accuracy for random forest: (a), (c) and (e), and sliding window accuracy with 5000 points (b), (d) and (f).

Table III shows results when using online methods for doing prediction on the three load traces. The accuracy performance is quite high for all traces. In some cases the online classifiers actually perform slightly better than the offline equivalents, even if the results are not significant. Hence the impact of incremental learning is minimal in these scenarios. Note that OAUE shows the best performance for all three traces.

Next in Table IV we present results from model training using random forest on one trace and predictions on another trace. This is equivalent to a concept drift where the concept changes rapidly and hence corresponds to the second research question. *The main takeaway is that doing predictions across traces introduce major challenges for offline methods.* As can be seen from Table IV, the classification accuracy degradation is quite noticeable in many cases. A similar behavior can be observed for the other metrics as well. Specifically, when learning from Periodic T1 and testing on Periodic T2 the classification accuracy is no better than random guessing. Similar performance behavior was observed using the two other offline methods.

Some of these results are also visualized in Figure 4 to further highlight the limitations of the offline approach. The figures show the asymptotic and temporal classification accuracy obtained by training a random forest model on one trace and then evaluating the model on another trace in a streaming fashion. Low asymptotic classification accuracy is evident in the asymptotic plots while high variability is visible in the temporal plots.

The key question, to be discussed next, is whether online learning can do better. Figure 5 shows results from experimenting with OAUE (best classifier based on results in Table III) where two different load traces are concatenated into a single long load trace. Three traces are constructed: *{Periodic T1, Periodic T2}*, *{Periodic T2, Periodic T1}* and *{Flash crowd, Periodic T2}*. The points where a trace ends and another starts is marked with a vertical line in the plots.

For the three concatenated load traces OAUE shows high performance not only on the first part of the stream but it continues to show consistently high performance on the second part of the stream as well. This was certainly not the case for offline methods. It is also evident that OAUE can maintain high classification accuracy despite concept drifts introduced by the new load traces. It should also be noted that the variations in sliding window performance are lower than what was observed using random forest (shown in Figure 4).

## VII. DISCUSSION AND CONCLUSIONS

Online machine learning models can learn *on-the-fly* in a streaming fashion and evolve their understanding of the system behavior when presented with new examples. This is a great advantage since for many applications the required data samples may not yet exist or they may arrive continuously as a stream. Hence, the learning model must gradually evolve with new samples.

In this paper we describe an application agnostic approach for real-time prediction of SLA violations, experienced on the client side, based solely on device statistics analyzed at the server side. We evaluate this system using a video-on-demand service by extensive experimentation and creation of load traces. These traces are used for online and offline model training and prediction in presence of concept drift.

The evaluations show that (1) offline methods can be used for accurate prediction for a specific load trace. Prediction across load traces is however not reliable. (2) Online methods provide high classification accuracy not only for a specific load trace but also across traces. That is, the results in this paper show that it is possible to predict SLA violations with high accuracy even under non-steady conditions like continuously changing load patterns.

Furthermore, we found that the online methods gradient ascent logistic regression and Hoeffding Tree provide accurate classification, both for a single load trace but also across multiple load traces. However, the best method for the current use case and load scenarios considered in this paper is OAUE.

The results reported in this paper are based on measurements from a relatively simple testbed configuration. But we are currently performing experiments with a distributed VoD service on a backend cluster [5]. Our goal is to extend and validate the results in this paper for a multi-server, multi service cloud environment paving the way to engineer management functions that predict SLA violations in real time with high accuracy under dynamic conditions.

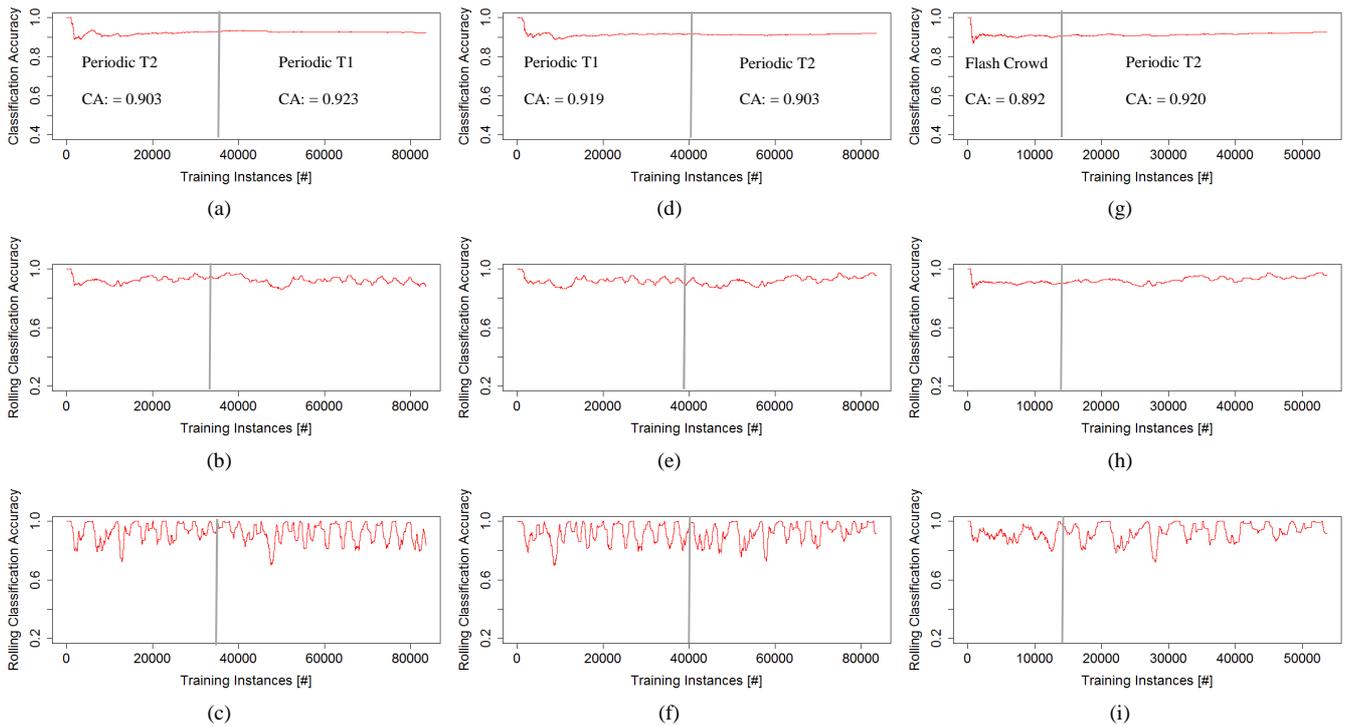

Fig. 5. Asymptotic classification accuracy for OAUE: (a), (d) and (g), sliding window accuracy with 5000 points (b), (e) and (h), sliding window accuracy with 1000 points window (c), (f) and (i).

ACKNOWLEDGMENT

This research has been supported by the Swedish Governmental Agency for Innovation Systems, VINNOVA, under grant 2013-03895.

REFERENCES

[1] Rerngvit Yanggratoke, Jawwad Ahmed, John Ardelius, Christofer Flinta, Andreas Johnsson, Daniel Gillblad, Rolf Stadler, "Predicting Real-Time Service-Level Metrics from Device Statistics", In IFIP/IEEE Integrated Management Conference (IM), Canada 2015.

[2] S. Handurukande, S. Fedor, S. Wallin, and M. Zach, "Magneto approach to qos monitoring," in Integrated Network Management (IM), 2011, IFIP/IEEE International Symposium on. IEEE, 2011, pp. 209–216.

[3] Song, Han Hee, et al. "Q-score: Proactive service quality assessment in a large IPTV system." Proceedings of the 2011 ACM SIGCOMM conference on Internet measurement conference. ACM, 2011.

[4] Sharma, Bikash, et al. "Cloudpd: Problem determination and diagnosis in shared dynamic clouds." Dependable Systems and Networks (DSN), 2013 43rd Annual IEEE/IFIP International Conference on. IEEE, 2013.

[5] Rerngvit Yanggratoke, Jawwad Ahmed, John Ardelius, Christofer Flinta, Andreas Johnsson, Daniel Gillblad, Rolf Stadler, A Platform for Predicting Real-Time Service-Level Metrics from Device Statistics, In IFIP/IEEE Integrated Management Conference (IM), Canada 2015.

[6] M. Shanmugavelayutham, "Data streams: Algorithms and applications", Now Publishers Inc, 2005.

[7] T. Hastie, R. Tibshirani, J. Friedman, and R. Tibshirani, "The elements of statistical learning", Vol. 2, no. 1. New York: springer, 2009.

[8] J. Gama, I. Žliobaitė, A. Bifet, M. Pechenizkiy, and A. Bouchachia. "A survey on concept drift adaptation." ACM Computing Surveys (CSUR) 46, no. 4 (2014): 44.

[9] J. Ross Quinlan, "C4.5: programs for machine learning", Morgan Kaufmann, San Francisco, 1993.

[10] L. Breiman, J. Friedman, R. Olshen, and C. J. Stone, "Classification and Regression Trees", Wadsworth and Brooks, Monterey, CA, 1984.

[11] P. Domingos, G. Hulten, "Mining high-speed data streams", Conference on Knowledge Discovery and Data Mining, pages 71–80, 2000.

[12] D. Brzezinski, and J. Stefanowski. "Combining block-based and online methods in learning ensembles from concept drifting data streams." Information Sciences 265 (2014): 50-67.

[13] Bifet, Albert, and Richard Kirkby. "Data stream mining a practical approach." (2009).

[14] Bifet, Albert, et al. "Efficient data stream classification via probabilistic adaptive windows." Proceedings of the 28th Annual ACM Symposium on Applied Computing. ACM, 2013.

[15] S. Godard, "SAR," http://linux.die.net/man/1/sar.

[16] Ceilometer: (online) http://docs.openstack.org/developer/ceilometer/

[17] VLC, (Online) http://www.videolan.org/vlc.

[18] D. Mills et al, "Network Time Protocol Version 4: Protocol and Algorithms Specification", Internet Engineering Task Force (IETF) RFC 5905, June 2010.

[19] R. Yanggratoke, J. Ahmed, J. Ardelius, C. Flinta, A. Johnsson, D. Gillblad, and R. Stadler, "Linux kernel statistics from a video server and service metrics from a video client." 2014, distributed by Machine learning data set repository [MLData.org]. http://mldata.org/repository/data/viewslug/realm-im2015-vod-traces.

[20] I. Ari, B. Hong, E. Miller, S. Brandt, and D. D. E. Long, "Managing flash crowds on the internet," in Modeling, Analysis and Simulation of Computer Telecommunications Systems, 2003. MASCOTS 2003. 11th IEEE/ACM International Symposium on, Oct 2003, pp. 246–249.

[21] RMOA: (Online) https://github.com/jwijffels/RMOA.

[22] MOA (Massive Online Analysis): (Online) http://moa.cms.waikato.ac.nz/.